\documentclass[runningheads]{llncs}
\usepackage{amsfonts}

\usepackage{latexsym}
\usepackage{amssymb}
\usepackage{epsfig}
\usepackage{amsmath}
\usepackage[mathscr]{eucal}
\usepackage{subfigure}
\usepackage{graphicx}

\newtheorem{Lemma}{Lemma}

\newtheorem{Theorem}[Lemma]{Theorem}

\renewcommand{\qed}{\hfill{\ \ \rule{2mm}{2mm}} \vspace{0.2in}}

\begin{document}

\title{Computing Prices for Target Profits in Contracts}
\author{ \textbf{Ghurumuruhan Ganesan}}
\authorrunning{G. Ganesan}
\institute{Institute of Mathematical Sciences, HBNI, Chennai\\
\email{gganesan82@gmail.com }}
\date{}
\maketitle

\begin{abstract}
Price discrimination for maximizing expected profit is a well-studied concept in economics and there are various methods that achieve the maximum given the user type distribution and the budget constraints. In many applications, particularly with regards to engineering and computing, it is often the case than the user type distribution is unknown or not accurately known. In this paper, we therefore propose and study a mathematical framework for price discrimination with \emph{target} profits under the contract-theoretic model. We first consider service providers
with a given user type profile and determine sufficient conditions for achieving a target profit.
Our proof is constructive in that it also provides a method to compute the quality-price tag menu. Next we consider a dual scenario where the offered service qualities
are predetermined and describe an iterative method to obtain nominal demand values that best match
the qualities offered by the service provider while achieving
a target profit-user satisfaction margin. We also illustrate our methods with design examples
in both cases.
\end{abstract}

\vspace{0.1in} \noindent \textbf{Key words:} Price discrimination, target profits, computational methods.

\setcounter{equation}{0}
\renewcommand\theequation{\arabic{section}.\arabic{equation}}
\section{Introduction}\label{intro}


Pricing of a product with different qualities is an important topic that has been studied extensively in economics under various scenarios. To determine the price tags for the various qualities, the company making the product usually performs extensive market research and creates a segmentation of the market into distinct groups, depending on the number of qualities available. Using a contract-theoretic model~\cite{mart}, it is then possible to provide enough incentive for the customers to buy the best quality product within their budget constraints.

The three main issues in any segmentation problem are service quality, user demand and the corresponding price tag. In this paper, we are interested in determining segmentations that achieve a target profit. There are many reasons for such a consideration. For example, suppose the user type distribution is not exactly known \emph{a priori} and therefore it might not be feasible to compute the true expected profit. This happens in practice when a new technology comes into practice and companies providing the technology do not have sufficient background data regarding user preferences in order to estimate the user type distribution and thereby maximize the overall expected profit.

Another scenario where target profit may be useful, is when the user types providing the largest profit to the service provider, occur with very low frequency. Thus while the expected profit may be large for the service provider, the actual (random) profit obtained from a randomly chosen user type might itself be low, with large probability. Ensuring a (minimum) target profit helps offset such vagaries.

\subsection*{Related Work}
In economics,~\cite{mussa} studied the multiproduct monopolist problem where a seller wants to provide a certain service at various qualities to cater to users of different types. The user type distribution has a density and the monopolist seeks to maximize the overall expected profit. Later in~\cite{arm} price discrimination for a many-product firm for customers with unobservable tastes (actions) is investigated. The goal there is to determine two-part tariffs that maximize the profits for the seller by extracting as much consumer surplus as possible. The customer actions are modelled as random variables and the tariff chosen maximizes the expected profit when the user action distribution is known. A related continuous parameter version of the problem is considered in~\cite{rochet} who introduce a sweeping operator in the context of multidimensional monopolist problem.

Similar economic issues arise in technological products as well. For example,~\cite{gao} uses contract-theory to study the problem of pricing the available qualities of transmission channels for the secondary (cognitive) users satisfying predetermined satisfaction levels . The type of a user is determined by its distance from the base station and the goal is to determine the prices for a given set of qualities and types of users (see also the book~\cite{gao2} for more details). Another example is the problem of spectrum trading in cognitive radio networks where primary users \emph{lease} spectrum to the secondary cognitive users with strict constraints regarding the time of usage and transmission power level. An adaptive learning algorithm is used in~\cite{niya} to estimate the instantaneous price of the spectrum and adjust individual demands accordingly.

Cloud and utility computing services is yet another area where pricing pays an important role with regards to usage of computing resources. In~\cite{yeo}, the pros and cons of charging fixed prices for metered usage of the services is analyzed and it is demonstrated via implementations that a variable pricing system which adjusts rates according to demand and quality of service requirements (QoS) is more beneficial in terms of revenue for the service. Finally, we remark that recently~\cite{wang} has studied resource allocation for network slices in 5G systems via network pricing and~\cite{xiong} investigate a multi-dimensional contract approach for data rewarding in mobile networks.

\subsection*{Our contributions}
To achieve target profits, we consider two possible cases: In the first case, the user type profile is fixed and we create a quality-price segmentation that achieves the target profit for the provider. In Theorem~\ref{thm1_pro} stated in Section~\ref{sec2}, we determine conditions on the cost and the user budget functions that ensure that a target profit margin~\(B\) is achieved. Our proof also provides a computational method to determine the service qualities offered and the corresponding price tags that best match the customer budget.

In the second case (Theorem~\ref{thm1}, Section~\ref{sec3}), the service quality profile is fixed and we describe an iterative computational procedure to obtain the corresponding nominal demand values that achieve the desired profit. We then set mutually beneficial prices for each of these nominal values so that any user can afford the nominal service quality closest to its own demand. We also study tradeoffs between the achievable profit for the service provider and the customer satisfaction margin.

The paper is organized as follows: In Section~\ref{sec2} we study achievable target profits for a given user demand profile and in Section~\ref{sec3} we describe conditions for achieving a target profit for a given service quality profile while also ensuring customer satisfaction.

\setcounter{equation}{0}
\renewcommand\theequation{\arabic{section}.\arabic{equation}}
\section{Target profits with given demand profile}\label{sec2}
A certain service provider wishes to offer services at~\(L\) different qualities to cater to~\(L\) different types of buyers with the aim of achieving a target profit. It costs~\(C(s)\) to offer service at quality~\(s > 0\) and users of type~\(i, 1 \leq i \leq L\) are willing to pay at most~\(P_i(s)\) for this service quality. The service provider would like to make a profit of~\(B(s)\) for offering service at quality~\(s.\) Typically~\(B(s)\) is a fraction of~\(C(s)\) and we say that the profit margin~\(B = B(.)\) is \emph{achievable} if there are quality values~\(0 < s_1 < s_2< \ldots <s_L \) and corresponding price tags~\(0 < p_1< \ldots<p_L\) satisfying the following for each~\(1 \leq k \leq L:\)
\begin{eqnarray}\label{ir_cond}
(x)&&P_k(s_k) \geq p_k \geq C(s_k)+B(s_k) \text{ and} \nonumber\\
(y)&&P_k(s_k) - p_k \geq P_k(s_l) - p_l \text{ for all }1 \leq l \neq k \leq L. \nonumber\\
\end{eqnarray}
We refer to condition~\((x)\) as the modified \emph{individual rationality} (IR) constraint that ensures that the price tag for service quality~\(s_k\) lies within the budget constraint of the buyer type~\(k,\) while also making a target profit~\(B(s_k)\) for the service provider. If~\(B(s_k) = 0,\) then condition~\((x)\) is simply called the individual rationality constraint~\cite{mart} and ensures that the service provider does not make a loss while providing the service at the given quality. Condition~\((y)\) is called the \emph{incentive compatibility} (IC) constraint that provides the necessary incentive (in terms of savings) for the buyer of type~\(k\) to actually buy service at quality~\(s_k\) (among all the qualities offered).

The following result determines conditions under the target profit~\(B\) is achievable.  Throughout~\(f'(.)\)
denotes the derivative of a function~\(f(.).\)
\begin{Theorem}\label{thm1_pro}
Suppose the functions~\(\{P_i(.)\}\) and~\(C(.)\) satisfy the following conditions:\\
\((a1)\) The cost and profit functions~\(C(s)\) and~\(B(s)\) are differentiable, strictly increasing and convex in~\(s\) with~\(C(0) =B(0) = 0.\)\\
\((a2)\) For~\(1 \leq i \leq L,\) the price function~\(P_i(s)\) is differentiable, strictly increasing and strictly concave in~\(s\) with~\(P_i(0)=  0.\) Moreover for~\(i < j,\) the derivatives~\(P'_i(s) < P'_j(s)\) for all~\(s > 0.\)\\
\((a3)\) There exists~\(x_1,y_L > 0\) such that~\(P_1(x_1) \geq C(x_1)+B(x_1)\) and~\(P_L(y_L) < C(y_L)+B(y_L).\)

If conditions~\((a1)-(a3)\) hold, then the target profit~\(B = B(.)\) is achievable.
\end{Theorem}
Conditions~\((a1)-(a2)\) in Theorem~\ref{thm1_pro} are usually true in practice since we expect that both the cost and the allowed budget to increase with the service quality. Condition~\((a2)\) is also called the \emph{single crossing condition}~\cite{varian}. Condition~\((a3)\) in Theorem~\ref{thm1_pro} ensures that the buyer with the smallest type finds a service quality within its budget and even the buyer with the largest type cannot afford very high quality services.

\subsection*{Proof of Theorem~\ref{thm1_pro}}
For~\(1 \leq i \leq L\) the function~\[f_i(s) := P_i(s)-C(s)-B(s)\]
is differentiable and strictly concave with~\(f_i(0)~=~0.\) Moreover, the set~\[A_i := \{f_i \geq 0\}\] is a closed and bounded (i.e., compact) interval
containing at least two points: Indeed, using condition~\((a2),\) we have for~\(i > j\) that~\[P_i(s) = \int_{0}^{s} P'_i(y) dy > \int_0^{s} P'_j(y) dy = P_j(s)\] and so~\[A_1 \subset A_2 \subset \ldots \subset A_L.\] Using~\(P_1(0) = 0=C(0)+B(0)\)
and~\(P_1(x_1) \geq C(x_1)+B(x_1)\) (condition~\((a3)\)) we then get that~\(\{0,x_1\} \subseteq A_1.\)

If~\(x,y \in A_i,\) then using concavity of~\(f_i,\)
we have for any~\(0 < \lambda < 1\) that~\[f_i(\lambda x + (1-\lambda)y) \geq \lambda f_i(x) + (1-\lambda)f_i(y) \geq 0.\]
Thus~\(A_i\) is an interval and using continuity of~\(f_i,\)
we have that~\(A_i\) is closed. The boundedness of~\(A_L\) follows from the fact that~\(f_L(y_L) <0\) (condition~\((a3)\)).

We now set~\(s_i > 0\) to be such that
\begin{equation}\label{genf}
f_i(s_i) = \max_{s \in A_i} f_i(s) =P_i(s_i)-C(s_i) -B(s_i) \geq  0
\end{equation}
and let~\(p_i:=C(s_i)+B(s_i).\) The derivative~\(f'_i\) is strictly decreasing and so~\(s_i\) is uniquely defined, since it must satisfy~\(f'_i(s_i) = 0.\)
Also, by condition~\((a2)\) we have that~\[f'_2(s_2) = 0 = f'_1(s_1) < f'_2(s_1)\] and so~\(s_1 < s_2.\) This also implies that
\[p_1 = C(s_1) +B(s_1) < C(s_2)+B(s_2) =  p_2\] since both~\(C(.)\) and~\(B(.)\) are strictly increasing (condition~\((a1)\)). Continuing iteratively we obtain
that
\begin{equation}\nonumber
0 < s_1 < s_2 < \ldots < s_L \text{ and }0 < p_1 < p_2 <\ldots <p_L.
\end{equation}
Moreover, from~(\ref{genf}) we get that~\(P_i(s_i) \geq p_i = C(s_i)+B(s_i)\)
so that condition~\((x)\) in~(\ref{ir_cond}) is true  and again using~(\ref{genf}), we get~\[P_i(s_i)-p_i = f_i(s_i) > f_i(s_j) = P_i(s_j)-p_j\] for all~\(1 \leq i \neq j \leq L.\) This implies that condition~\((y)\) in~(\ref{ir_cond}) is also true.~\(\qed\)

We now illustrate the above proof with an example.
\subsection*{Example}
Suppose the cost for providing service at quality~\(s\) increases linearly with~\(s\) and the service provider sets a target profit of~\(10\%.\)
In our notation we set~\[C(s) = D_c \cdot s \text{ and } B(s) = \frac{C(s)}{10} = D_c \cdot \frac{s}{10}\] be the cost and profit functions, respectively, for some constant~\(D_c > 0.\) Also suppose that there are~\(L\) types of users and the user budget increases \emph{logarithmically} in~\(s\) and is proportional to the type number. Formally, we set~\[P_i(s) = D_b \cdot i\cdot \log(1+s)\] for some positive constant~\(D_b.\)

With the above definitions, conditions~\((a1)-(a2)\) in Theorem~\ref{thm1_pro} are satisfied. Moreover, if~\(D_b > \frac{11D_c}{10}\) strictly, then the condition~\((a3)\) in Theorem~\ref{thm1_pro} is true as well, since~\[P_1(s) = D_b \cdot \log(1+s) \geq \frac{11\cdot D_c}{10} \cdot s = C(s) + B(s)\] for all small~\(s > 0\) and~\[P_L(s) = D_b \cdot L \cdot \log(1+s) <  \frac{11D_c s}{10} = C(s) + B(s)\] for all large~\(s.\) From~(\ref{genf}), we therefore get that the quality~\(s_i\) that achieves the target profit for the service must satisfy~\(f_i(s_i) = \max_{s} f_i(s)\) where\[f_i(s) := D_b\cdot i \cdot \log(1+s) - \frac{11D_cs}{10}.\] Differentiating~\(f_i(s)\) and equating to zero, we get that~\(s_i = \frac{10D_b}{11D_c}\cdot i-1.\) Thus the service quality offered to a particular user varies linearly with its type.



\setcounter{equation}{0}
\renewcommand\theequation{\arabic{section}.\arabic{equation}}
\section{Target profits with given quality profile}\label{sec3}
Consider a service provider who already offers services at predetermined qualities and would like to assign price tags to these qualities based on the user demands. There are~\(L\) different qualities~\(s_{low} \leq s_1 < s_2 < \ldots s_{L-1} < s_L \leq s_{up}\) and users with different demands request corresponding qualities from the service provider, taking into account the budget constraint. We assume that all user demands lie within an interval~\([\theta_{low},\theta_{up}]\) in the positive real line. A user with demand~\(\theta \in [\theta_{low},\theta_{up}]\) is willing to pay up to~\(F(\theta,s)\) for service quality~\(s,\) while it costs~\(C(s)\) for the service provider to deliver at such quality.

It is of interest to determine nominal demand values~\[\theta_{low} \leq \theta_1 < \theta_2 < \ldots \theta_L \leq \theta_{up}\] tagged with corresponding prices~\(p_1 < p_2 < \ldots <p_L\) so that any user with demand~\(\theta \in [\theta_k,\theta_{k+1})\) is able to pay price~\(p_k\) and utilize service at quality~\(s_k.\) If the demand~\(\theta\) exactly equals~\(\theta_k,\) then the user also is able to save much more by choosing quality~\(s_k\) than any of the other qualities. Formally, we prefer that conditions~\((x)-(y)\) stated in~(\ref{ir_cond}) hold with~\(P_k(s_k)=F(\theta_k,s_k).\)

One bottleneck in the above formulation is that users with demands not equal to one of the nominal values are not guaranteed savings though they can still afford the services. It might possibly result in few number of users (those with demands exactly equal to one of the nominal values) being satisfied in terms of savings. In addition, the service provider itself may have a target profit in mind. Taking these into account, we have the following modified definition.


Let~\(\mathbf{b} = (b_1,\ldots,b_L), \mathbf{m} = (m_1,\ldots,m_L)\) be any two~\(L-\)tuples satisfying~\[0 < b_i < b_{i+1}, 0< m_i < m_{i+1}, 1 \leq i \leq L-1.\] The service provider is said to achieve a profit-satisfaction margin~\((\mathbf{b},\mathbf{m})\) if there exists a demand-price profile~\(\{(\theta_k,p_k)\}_{1 \leq k \leq L}\) satisfying~\(\theta_{low}\leq \theta_1 < \theta_2 < \ldots < \theta_L \leq \theta_{up}\) and~\(0 < p_1 < p_2 < \ldots <p_L\) such that the following holds: For any~\((\hat{\theta}_1,\ldots,\hat{\theta}_L)\) satisfying~\(|\hat{\theta}_k-\theta_k| \leq m_k,1 \leq k \leq L\) we have:
\begin{eqnarray}
&(i)& F(\hat{\theta}_k,s_k) \geq p_k \geq C(s_k)+b_k \text{ (IR) and }\nonumber\\
&(ii)& F(\hat{\theta}_k,s_k) -p_k \geq F(\hat{\theta}_k,s_l) -p_l\;\;\forall\;\;l \neq k \text{ (IC)}.\;\;\;\;\;\label{ir_cond2}
\end{eqnarray}
Essentially conditions~\((i)-(ii)\) ensure that users with demand in the range\\\([\theta_k-m_k,\theta_k +m_k]\) make savings if they choose service at quality~\(s_k.\) At the same time, the service provider also secures a profit of~\(b_k\) for providing the service at quality~\(s_k.\) The quantities~\(\{\theta_k\}\) are the nominal demand values determined by the service provider that results in its net profit. 

Throughout we assume that the function~\(F(\theta,s)\) has support in the compact interval~\([\theta_{low},\theta_{up}] \times [ s_{low},s_{up}],\) and that both its partial derivatives are positive, increasing and jointly continuous in both~\(\theta\) and~\(s.\) We also assume that~\[\frac{\partial^2 F(\theta,s)}{\partial \theta \partial s} = \frac{\partial^2 F(\theta,s)}{\partial s \partial \theta} > 0\] is jointly continuous in both~\(\theta\) and~\(s\) and for simplicity denote~\[F_s(\theta,s) := \frac{\partial F(\theta,s)}{\partial s}, F_{\theta}(\theta,s) := \frac{\partial F(\theta,s)}{\partial \theta} \text{ and } F_2(\theta,s) := \frac{\partial^2 F(\theta,s)}{\partial s \partial \theta}.\] For~\(2 \leq j \leq L,\) we define
\begin{equation}\label{del_def}
\epsilon_j := \sup_{\theta} F_{\theta}(\theta,s_{j-1}) \text{ and }\delta_j := \inf_{\theta} \left(F_{\theta}(\theta,s_{j})- F_{\theta}(\theta,s_{j-1})\right),
\end{equation}
where the infimum and supremum are taken over all~\(\theta \in [\theta_{low},\theta_{up}]\) and have the following result regarding the achievability of a target profit-satisfaction margin.
\begin{Theorem}\label{thm1}
Suppose that
\begin{equation}\label{imp_cond}
\inf_{\theta} F_s(\theta,s) \geq \frac{dC(s)}{ds} \;\;\;\forall s
\end{equation}
where the infimum and supremum are taken over all~\(\theta \in [\theta_{low},\theta_{up}].\) If the profit-satisfaction margin~\((\mathbf{b},\mathbf{m})\) satisfies~\[\sum_{j=1}^{L}\Delta_{j} + m_L < \theta_{up}-\theta_{low}\] and~\(P(\theta_{low},s_1) \geq C(s_1)+ b_1,\) where~\(\Delta_1 = m_1\) and
\begin{equation}\label{del_def2}
\Delta_{j} := (m_j + m_{j-1})\left(1 + \frac{2\epsilon_j}{\delta_j}\right) + \frac{b_j-b_{j-1}}{\delta_j}
\end{equation}
for~\(2 \leq j \leq L,\) then~\((\mathbf{b},\mathbf{m})\) is achievable.
\end{Theorem}
The condition in~(\ref{imp_cond}) (which we call the marginal budget increase condition) essentially says that the marginal budget increase for any user type is at least as large as the marginal cost increase. We obtain the desired profit-satisfaction margin by introducing an additional \emph{profit constraint} that provides minimum incentives for users with demand~\(\theta_k\) to select service quality~\(s_k\) than at a slightly better quality~\(s_{k+1}.\) We then use an induction procedure analogous to~\cite{gao} to demonstrate the achievability.

\subsection*{Example}
For~\(\delta > 0\) and~\(0 < b,m < 1\) suppose~\[s_i = i \cdot \delta, b_i=b \cdot s_i \text{ and }m_i = m \cdot s_i\] for~\(1 \leq i \leq L\) so that the service would like to achieve a flat profit of~\((100b) \%.\) The cost of providing service~\(C(s) = D_c \cdot s\) for some constant~\(D_c > 0\)
and we assume that the budget~\(F(\theta,s) = D_p \cdot \theta \cdot s\) where~\(D_p \geq \theta_{low}^{-1} D_c\) so that condition~(\ref{imp_cond}) is true.

From~(\ref{del_def}), we have for~\(2 \leq j \leq L\) that~\(\delta_j = D_p\cdot \delta\) and~\(\epsilon_j=D_p (j-1)\delta\) so and substituting into~(\ref{del_def}) we get that~\(\Delta_1 = ms_1 = m\delta\) and~\(\Delta_{j} = m\delta(2j-1)^2 + \frac{b}{D_p}\) for~\(2 \leq j \leq L.\) Letting~\(\Delta(S) := s_{up}-s_{low}\) be the service quality  range and~\(\Delta(\theta) := \theta_{up}-\theta_{low}\) denote the demand range, we then get
\[\sum_{j=1}^{L}\Delta_{j} +m_L <  4m\delta\sum_{j=1}^{L} j^2 + mL\delta < 4m\delta L^3 = 4m\Delta(S) L^2.\] Therefore from Theorem~\ref{thm1}, we get that the set of achievable profit-customer satisfaction margins is given by
\begin{eqnarray}\label{eq_hom}
{\cal A} &:=& \left\{ (b,m) : b > 0, 0 < m < 1 \text{ and }\right. \nonumber\\
&&\;\;\;\;\;\left.m\cdot\left(4\Delta(S) L^2\right) + b\cdot\left(\frac{L}{D_p}\right) \leq \Delta(\theta)\right\}.
\end{eqnarray}

The expression in~(\ref{eq_hom}) also determines the tradeoff between profit and customer satisfaction: For a given range of user demands, higher customer satisfaction can be obtained at the cost of reduced profit.
For zero profit \(b =0,\) the maximum achievable customer satisfaction margin is~\(m_0 := \min\left(1,\frac{\Delta(\theta)}{4 \Delta(S) L^2}\right)\)
and grows as the square of the inverse of the number of types~\(L.\) Similarly for zero customer satisfaction margin~\(m=0,\) the maximum achievable profit is~\(b_0 := \frac{\Delta(\theta)}{D_pL}\) which grows inversely as the number of types.

\begin{figure}[tbp]
\centering
\includegraphics[width=2.5in]{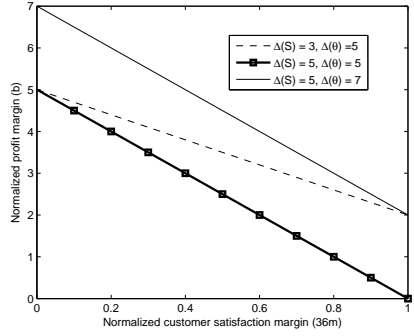}
\caption{Tradeoff curves for various values of~\(\Delta(S)\) and~\(\Delta(\theta).\)}
\label{trade_off}
\end{figure}

To illustrate the tradeoff in~(\ref{eq_hom}), we consider the case of~\(L = 3\) service qualities with~\(D_c = 1\) and~\(D_p = \frac{1}{\theta_{low}} = 3.\)
From~(\ref{eq_hom}) we then have that the tradeoff curve is~\[36 m \cdot \Delta(S) + b =\Delta(\theta)\]
and in Fig.~\ref{trade_off}, we plot the tradeoff curves as a function of the normalized customer satisfaction margin~\(36m\) and the normalized profit margin~\(b\) for various values of~\(\Delta(S)\) and~\(\Delta(\theta).\)
From the solid and dashed lines in Fig.~\ref{trade_off}, we see that if the quality range~\(\Delta(S)\) is fixed, then any increase in the demand range~\(\Delta(\theta)\) results in increased profit for the same customer satisfaction margin. On the other hand, from the solid line and the marked line, we infer that increasing the quality range while the demand range remains fixed, might not be profitable.

\subsection*{Proof of Theorem~\ref{thm1}}
We prove a slightly stronger result. For~\(1 \leq k \leq L-1,\) let~\(f_k(\mathbf{b}) \geq b_{k+1}-b_k \geq 0\) be functions such that
\begin{equation}\label{p_cond}
F(\theta_{low},s_{1}) \geq C(s_{1}) + b_1 \text{ and }\sum_{j=1}^{L} \Delta_{j} < \theta_{low}-\theta_{up},
\end{equation}
where~\(\Delta_1 = m_1\) and for~\(2 \leq j \leq L,\)
\begin{equation}\label{del_def3}
\Delta_{j} := (m_j + m_{j-1})\left(1 + \frac{2\epsilon_j}{\delta_j}\right) + \frac{f_{j-1}(\mathbf{b})}{\delta_j}
\end{equation}
and~\(\epsilon_j\) and~\(\delta_j\) are as in~(\ref{del_def}). Letting~\(\omega_k := F(\theta_{k}-m_k,s_{k}) -p_{k},\) we prove that there are constants~\(\{(\theta_k,p_k)\}_{1 \leq k \leq L}\) such that for each~\(1 \leq k,l \leq L:\)
\begin{eqnarray}
&(a)& \omega_k \geq 0  \text{ and } p_k \geq C(q_k)+b_k. \nonumber\\
&(b)& \omega_k \geq f_{k}(\mathbf{b}) + F(\theta_{k}+m_k,s_{k+1}) -p_{k+1}. \nonumber\\
&(c)& \omega_k \geq F(\theta_k+m_k,s_l) -p_l, l \neq k. \label{ir_cond3}
\end{eqnarray}
Condition~\((b)\) is an extra condition which we impose and call as the \emph{profit constraint}. The profit constraint ensures that extra incentive is provided for user with demand~\(\theta_k\) to use service quality~\(s_k\) than a slightly better quality~\(s_{k+1}.\) Conditions~\((a)\) and~\((c)\) are modifications of the individual rationality constraint and the incentive compatibility constraints, respectively, discussed in Section~\ref{sec2}. Because~\(F(\theta,s)\) is increasing in~\(\theta,\) conditions~\((a)\) and~\((c)\) respectively ensure that conditions~\((i)\) and~\((ii)\) in~(\ref{ir_cond2}) are satisfied.

Analogous to~\cite{gao}, we perform the desired pricing iteratively.  Let~\(\theta_1 := \theta_{low}+m_1\) and let the price~\(p_1\) be such that~\(\omega_1 \geq p_1 \geq C(s_1)+b_1.\) This is possible by the statement of Theorem~\ref{thm1}. Suppose now conditions~\((a)-(c)\) in~(\ref{ir_cond3}) are true with~\(L\) replaced by~\(j-1\) for some~\(2 \leq j \leq L-1.\) We determine~\((\theta_{j},p_j)\) as follows. If~\((a)-(c)\) were to be satisfied with~\(L\) replaced by~\(j,\) then from the profit constraint~\((b)\) we must have
\begin{eqnarray}
p_j &\geq& p_{j-1} + F(\theta_{j-1}+m_{j-1},s_j) -\omega_{j-1} + f_{j-1}(\mathbf{b}) \nonumber\\
&=& p_{j-1} + \int_{s_{j-1}}^{s_j} F_s(\theta_{j-1}+ m_{j-1},s) \nonumber\\
&&\;\;\;\;\;\;\;+\;\;\int_{\theta_{j-1}-m_{j-1}}^{\theta_{j-1}+m_{j-1}} F_{\theta}(\theta,s_{j-1}) + f_{j-1}(\mathbf{b}) \nonumber\\
&=:&  A_j.\label{eqq1}
\end{eqnarray}
Similarly from IC constraint~\((c)\) we must have
\begin{eqnarray}
p_j &\leq& p_{j-1} + \omega_j - F(\theta_{j}+m_j,s_{j-1})  \nonumber\\
&=& p_{j-1} + \int_{s_{j-1}}^{s_j} F_s(\theta_{j}-m_j,s)- \int_{\theta_{j}-m_{j}}^{\theta_{j}+m_{j}} F_{\theta}(\theta,s_{j-1}) \nonumber\\
&=:& B_j. \label{eqq2}
\end{eqnarray}
Thus we need to choose~\(\theta_j\) such that~\(J_1-J_2 \geq f_{j-1}(\mathbf{b}),\) where~
\begin{eqnarray}
J_1 &:=& \int_{s_{j-1}}^{s_j} F_s(\theta_{j}-m_j,s) - \int_{s_{j-1}}^{s_j} F_s(\theta_{j-1}+ m_{j-1},s) \nonumber\\
&=& \int_{s_{j-1}}^{s_j}\int_{\theta_{j-1}+m_{j-1}}^{\theta_j-m_j} F_2(\theta,s) \nonumber\\
&=& \int_{\theta_{j-1}+m_{j-1}}^{\theta_j-m_j} F_{\theta}(\theta,s_j) - F_{\theta}(\theta,s_{j-1}) \nonumber
\end{eqnarray}
and~\[J_2 := \int_{\theta_{j}-m_{j}}^{\theta_{j}+m_{j}} F_{\theta}(\theta,s_{j-1}) + \int_{\theta_{j-1}-m_{j-1}}^{\theta_{j-1}+m_{j-1}} F_{\theta}(\theta,s_{j-1}).\] Recalling the definition of~\(\delta_j \) and~\(\epsilon_j\) from~(\ref{del_def}),
we know that~\[J_1-J_2 \geq \delta_j(\theta_j-\theta_{j-1}-(m_j+m_{j-1})) - \epsilon_j(2m_j + 2m_{j-1})\]
and so setting
\begin{equation}\label{thet_choice}
\theta_{j} := \theta_{j-1} + \Delta_{j}
\end{equation}
where~\(\Delta_j\) is as in~(\ref{del_def2}), we get the desired inequality~\(J_1 -J_2 \geq f_{j-1}(\mathbf{b}).\) From~(\ref{eqq1}) and~(\ref{eqq2}) we also get that~\(A_j \leq B_j\) and so we choose~\(p_j\) such that~\(A_j \leq p_j \leq B_j.\)

With this choice of~\((\theta_j,p_j)\) we now check if conditions~\((a)-(c)\) in~(\ref{ir_cond3})
are satisfied with~\(L\) replaced by~\(j\) and this would prove the induction step.
Using~(\ref{eqq1}) and the fact that~\(F_{\theta} > 0\) we get
\begin{equation}\label{eqt}
p_j \geq p_{j-1} + \int_{s_{j-1}}^{s_j} F_s(\theta_{j-1}+ m_{j-1},s)  + f_{j-1}(\mathbf{b}).
\end{equation}
By the marginal budget increase condition~(\ref{imp_cond}), we know
that~\[\int_{s_{j-1}}^{s_j} F_s(\theta_{j-1}+ m_{j-1},s) \geq C(s_j)-C(s_{j-1})\]
and so we get~\(p_j \geq C(s_j) + f_{j-1}(\mathbf{b}) + b_{j-1}\) since condition~\((a)\) in~(\ref{ir_cond3}) holds with~\(k = j-1.\) Using~\(f_{j-1}(\mathbf{b}) \geq {b}_j -{b}_{j-1},\) we then get that~\(p_j \geq C(s_j) + {b}_j,\) verifying the second condition of~\((a)\) with~\(k\) replaced by~\(j.\)
To verify the first condition of~\((a),\) we have from~(\ref{eqq2}) that
\begin{equation}
p_j\leq p_{j-1} + \omega_j - F(\theta_{j}+m_j,s_{j-1})  \leq p_{j-1} + \omega_j - \omega_{j-1} \nonumber
\end{equation}
since~\(F(\theta,s)\) is increasing in~\(\theta\) and~\(\theta_{j-1} -m_{j-1}< \theta_j+m_j.\)  Using condition~\((a)\) with~\(L=j-1,\) we have~\(p_{j-1} \leq  \omega_{j-1}\) and so~\(p_j \leq \omega_j.\) This verifies the first condition in~\((a)\) with~\(L\) replaced by~\(j.\)

By construction~\(A_j \geq B_j\) (see~(\ref{eqq1}) and~(\ref{eqq2})) and so the profit constraint~\((b)\) in~(\ref{ir_cond3}) is true with~\(k=j-1.\)
From~(\ref{eqq2}) we have that condition~\((c)\) in~(\ref{ir_cond3}) is true with~\(k\) replaced by~\(j\) and~\(l\) replaced by~\(j-1.\) Further using~\(\theta_{j-1}-m_{j-1} < \theta_j +m_j\) and the fact that~\(F(\theta,s)\) is increasing in~\(\theta,\)  we have for~\(1 \leq l \leq j-1\) that~\(\omega_j-p_j\) is bounded below by
\begin{equation}
F(\theta_{j}+m_{j},s_{j-1})-p_{j-1}\geq \omega_{j-1}-p_{j-1} \geq F(\theta_{l}+m_{l},s_{l})-p_{l}, \nonumber
\end{equation}
since by induction assumption, condition~\((c)\) in~(\ref{ir_cond3}) is true when~\(L\) and~\(k\) are replaced by~\(j-1\) and~\(l \leq j-1,\) respectively.

For the reverse direction, we again use induction to get for~\(1 \leq l \leq j-1\) that~\(\omega_l \geq F(\theta_{l}+m_l,s_{j-1})-p_{j-1}.\) Using~(\ref{eqt}) with~\(f_{j-1}(\mathbf{b}) \geq 0\) to estimate~\(p_{j-1},\) we get that~\(F(\theta_{l}+m_l,s_{l})-p_{l} \) is at least
\begin{eqnarray}
&&F(\theta_l+m_l,s_{j-1})-p_j + \int_{s_{j-1}}^{s_j} F_s(\theta_{j-1}+ m_{j-1},s) \nonumber\\
&&\;\;\geq\;\;F(\theta_l+m_l,s_{j-1})-p_j+\int_{s_{j-1}}^{s_j} F_s(\theta_{l}+ m_{l},s) \nonumber\\
&&\;\;=\;\;F(\theta_l+m_l,s_j)-p_j, \nonumber
\end{eqnarray}
since~\(F_s(\theta,s)\) is increasing in~\(\theta\) and~\(\theta_l + m_l < \theta_{j-1}+m_{j-1}.\) This verifies condition~\((c)\) in~(\ref{ir_cond3}) with~\(L\) replaced by~\(j,\) completing the induction step with~\(L\) replaced by~\(j \leq L-1.\)

Finally, we get from the discussion following~(\ref{thet_choice}) that~\[\sum_{j=1}^{L-1}\Delta_{j} \leq \theta_{up}-\theta_{low} - \Delta_{L}-m_L\]
by~(\ref{del_def3}). Thus~\(\theta_{up}-\theta_{L-1} \geq \Delta_{L}+m_L\) and setting~\(\theta_L := \theta_{L-1}+\Delta_L\) as in~(\ref{thet_choice}), we get that~\(\theta_L + m_L \leq \theta_{up}.\) Proceeding as in the discussion following~(\ref{thet_choice}), we then complete the induction step for~\(j=L.\)~\(\qed\)

\section{Conclusion and future work}
In this paper, we have described a computational framework for securing target profits for service providers under the contract-theoretic model for a given user type profile and a given service quality profile. We also illustrated our results with design examples. In the future, we plan to analyse further challenges that are unique to applications. As a typical example, consider a wireless network where some users are far from the base station and therefore may not have as good service quality as those close to the station. But if these far away users are willing to pay extra money, then how could the provider ensure high quality service for such users? It would be interesting to study the profit-satisfaction tradeoff under such situations.

\section*{Acknowledgement}
I thank Professors Rahul Roy, Arunava Sen, C. R. Subramanian and the referees for crucial comments that led to an improvement of the pape. I also thank IMSc for my fellowships.

\end{document}